\documentstyle[amssymb,12pt,epsfig,psfig]{article}

\makeatletter

\begin{document}

\title{ ON THE THEORY OF COHERENT PAIR PRODUCTION IN CRYSTALS IN PRESENCE
OF ACOUSTIC WAVES}
\author{A. R. Mkrtchyan, A. A. Saharian\footnote{%
E-mail: saharyan@server.physdep.r.am}, L. Sh. Grigoryan, B. V. Khachatryan\\
{\it Institute of Applied Problems in Physics,}\\ {\it 25
Nersessian Str., 375014 Yerevan, Armenia}}

\maketitle

\begin{abstract}
The influence of hypersonic waves excited in a single crystal is
investigated on the process of electron-positron pair creation by
high-energy photons. The coherent part of the corresponding
differential cross-section is derived as a function of the
amplitude and wave number of the hypersound. The values of the
parameters are specified for which the latter affects remarkably
on the pair creation cross-section. It is shown that under certain
conditions the presence of  hypersonic waves can result in
enhancement of the process cross-section.
\end{abstract}

\section{Introduction}

The fundamental aspects of the high-energy electromagnetic
processes in condensed media have been under discussion for a long
time. In crystals, where the atoms are arranged in ordered rows,
the cross-sections of the high-energy electromagnetic processes
can change essentially compared with the corresponding quantities
for a single atom (see, for instance,
\cite{TerMik,Shulga,Bazy87,Baie89} and references therein). The
cross-sections change because the kinematics of the processes
dictate that the interactions be spread over a significant
distance. The momentum transfer between a highly relativistic
interacting particle and the crystal can be small, especially
along the direction of particle motion. When this longitudinal
momentum transfer is small, the uncertainty principle dictates
that the interaction be spread out over a distance, known as the
formation length for particle production or, more generally, as
the coherence length. If the formation length exceeds the
interatomic spacing the interference effects from all atoms within
this length are important and they can essentially affect the
corresponding cross-sections.

From the point of view of controlling the parameters of the
high-energy electromagnetic processes in a medium it is of
interest to investigate the influence of external fields, such as
acoustic waves, temperature gradient etc., on the corresponding
characteristics. The considerations of concrete processes, such as
diffraction radiation \cite{MkrtDR}, transition radiation
\cite{Grigtrans}, parametric X-radiation \cite{Mkrt91}, channeling
radiation \cite{Mkrtch1},  have shown that the external fields can
essentially change the angular - frequency characteristics of the
radiation intensities. In the present paper we investigate the
influence of a hypersonic wave on the coherent creation of
electron-positron pairs by high-energy photons in a crystal (for
the analogous influence in the case of bremsstrahlung see
\cite{Sahbrem}). We specify the conditions under which the
external deformation field changes the pair creation cross-section
compared to the case of an undeformed crystal and demonstrate the
possibility for the positron yield enhancement. Electron-positron
pair creation phenomena by high-energy photons passing through a
crystal are interesting not only from the viewpoint of underlying
physics (for a recent review see \cite{Kuna00}) but also from the
viewpoint of practical applications for generation of intense
positron beams. The basic source to creating positrons for
high-energy electron-positron colliders is the electron-positron
pair creation by hard bremsstrahlung photons produced when a
powerful electron beam hits an amorphous target. Another approach
is to use a crystal, the atomic rows of which are aligned with the
incident electron beam, instead of an amorphous target. Due to the
coherent bremsstrahlung this gives more intense radiation of
photons, and as a result higher positron intensity. To further
improve the positron yield the channeling radiation can be used,
which is more intense than coherent bremsstrahlung and gives more
electron-positron pairs
\cite{Cheh89,Deck91,Artr94,Artr96,Yosh98,Kali98,Baie98,Inou01,CERN01}.
An additional increase in the yield of positrons can be reached by
using acoustic waves excited in crystals \cite{Wagn99}.

The paper is organized as follows. In the next section a formula
is derived for the coherent part of the electron-positron pair
creation by a photon in presence of the sinusoidal deformation
field generated by a hypersound. The analysis of the general
formula and numerical results in the special cases when the photon
enters into the crystal at small angle with respect to
crystallographic axes or planes is presented in Sec.
\ref{sec3:an}. The main results are re-mentioned and discussed in
Sec. \ref{sec4:conc}.

\section{Cross-section of the pair creation} \label{sec2:form}

The differential cross-section for the electron-positron pair
creation in a crystal by a high-energy photon with the wave vector
${\mathbf{k}}$ can be written in the form (see, for example,
\cite{TerMik,Shulga})
\begin{equation}
\sigma ({\bf q})\equiv \frac{d^{4}\sigma }{dE_{+}d^{3}q}=
\left| \sum_{n}e^{i{\bf qr}_{n}}\right| ^{2}\sigma _{0}(%
{\bf q}),  \label{sig1}
\end{equation}
with ${\mathbf{r}}_{n}$ being the positions of atoms in the
crystal, $\sigma _{0}({\mathbf{q}})$ is the corresponding
cross-section on an individual atom as a function of momentum
transfer to the crystal ${\bf q}={\bf k}-{\bf p}_{+}-{\bf p}_{-}$.
Here ${\bf p}_{+}$, $E_{+}$ and ${\bf p}_{-}$, $E_{-}$ are the
momenta and energies for the craeted positron and electron
respectively (hereafter the system of units $\hslash =c=1$ is
used). The interference factor in Eq. (\ref{sig1}) is responsible
for coherent effects arising due to periodical arrangement of the
atoms in a crystal. After averaging on thermal fluctuations the
cross-section takes the standard form \cite{TerMik}
\begin{equation}
\sigma ({\bf q})=\left\{ N_{0}\left( 1-e^{-q^{2}\overline{u_{t}^{2}}}\right)
+e^{-q^{2}\overline{u_{t}^{2}}}\left| \sum_{n}e^{i{\bf qr}_{n0}}\right|
^{2}\right\} \sigma _{0}({\bf q}),  \label{sig2}
\end{equation}
where ${\bf r}_{n}={\bf r}_{n0}+{\bf u}_{tn}$, ${\bf u}_{tn}$ is
the displacement of atoms with respect to the equilibrium
positions ${\mathbf{r}}_{n0}$ (by taking into account the crystal
deformation due to the hypersonic wave) due to the thermal
vibrations , $\overline{u_{t}^{2}}$ is the temperature dependent
mean-squared amplitude of the thermal vibrations of atoms in the
crystal, $N_{0}$ is number of atoms in the crystal,
$e^{-q^{2}\overline{u_{t}^{2}}}$ is the Debye-Waller factor. When
external influences are present the positions of atoms can be
written as
\begin{equation}
{\bf r}_{n0}={\bf r}_{ne}+{\bf u}_{n},  \label{rn0}
\end{equation}
with ${\bf r}_{ne}$ being the equilibrium positions of atoms in
the situation without deformation, ${\bf u}_{n}$ are the
displacements of atoms caused by the acoustic wave. We will
consider deformations with the sinusoidal structure
\begin{equation}
{\bf u}_{n}={\bf u}_{0}\sin \left( {\bf k}_{s}{\bf r}_{ne}+\varphi
_{0}\right) , \label{uacust}
\end{equation}
where ${\bf k}_{s}$ is the wave vector of the hypersonic wave.
Note that the dependence of ${\bf u}_{n}$ on time (through the
phase $\varphi _{0}$) we can disregard, as for particle energies
we are interested in, the characteristic time for the change of
deformation field is much greater than the passage time of
particles through the crystal. For the deformation field given by
Eq. (\ref{uacust}) the sum over the atoms of a crystal in
(\ref{sig2}) can be transformed into the form
\begin{equation}
\sum_{n}e^{i{\bf qr}_{n0}}=\sum_{m=-\infty }^{+\infty }J_{m}({\bf qu}%
_{0})e^{im\varphi _{0}}\sum_{n}e^{i{\bf q}_{m}{\bf r}_{ne}},\quad {\bf q}%
_{m}={\bf q}+m{\bf k}_{s},  \label{qm}
\end{equation}
where $J_{m}(x)$ is the Bessel function. For a lattice with a
complex cell the coordinates of the atoms can be written as
\begin{equation}
{\bf r}_{ne}={\bf R}_{n}+{\bf \rho }_{j},  \label{rne}
\end{equation}
with ${\bf R}_{n}$ being the positions of the atoms for one of
primitive lattices, and ${\mathbf{\rho }}_{j}$ are the equilibrium
positions for other atoms inside $n$ - th elementary cell with
respect to ${\bf R}_{n}$ . Now the sum over atoms of the lattice
can be presented as
\begin{equation}
\sum_{n}e^{i{\bf q}_{m}{\mathbf{r}}_{ne}}=S({\mathbf{q}}_{m})\sum_{n}e^{i{\mathbf{q}}_{m}%
{\mathbf{R}}_{n}},\quad
S({\mathbf{q}})=\sum_{j}e^{i{\mathbf{q}}{\mathbf{\rho }}_{j}},
\label{Sq}
\end{equation}
where $S({\mathbf{q}})$ is the structure factor. By making use of
the formulae given above the square of the modulus for the sum
(\ref{qm}) takes the form
\begin{equation}
\left| \sum_{n}e^{i{\bf qr}_{n0}}\right| ^{2}=\sum_{m,m^{\prime }=-\infty
}^{+\infty }J_{m}\left( {\bf qu}_{0}\right) J_{m^{\prime }}\left( {\bf qu}%
_{0}\right) e^{i(m-m^{\prime })\varphi _{0}}\sum_{n,n^{\prime }}e^{i{\bf q}%
_{m}{\bf R}_{n}}e^{-i{\bf q}_{m^{\prime }}{\bf R}_{n^{\prime }}}S({\bf q}%
_{m})S^{\ast }({\bf q}_{m^{\prime }}).  \label{SmSm}
\end{equation}
For thick crystals the sum over cells can be presented as a sum
over the reciprocal lattice:
\begin{equation}
\sum_{n}e^{i{\bf q}_{m}{\bf R}_{n}}=\frac{(2\pi )^{3}}{\Delta }\sum_{{\bf g}%
}\delta ({\bf q}_{m}-{\bf g}),  \label{sumdelta}
\end{equation}
where $\Delta $ is the unit cell volume, and ${\bf g}$ is the
reciprocal lattice vector. By taking into account the $\delta $ -
function the quantity ${\bf q} _{m^{\prime }}$ can be written as
${\bf q}_{m^{\prime }}={\bf g}+(m^{\prime
}-m){\bf k}_{s}$. As for the lattice one has $e^{-i{\bf gR}%
_{n}}=1$, we receive
\begin{equation}
\sum_{n^{\prime }}e^{-i{\bf q}_{m^{\prime }}{\bf R}_{n^{\prime
}}}=\sum_{n^{\prime }}e^{-i(m^{\prime }-m){\bf k}_{s}{\bf R}_{n^{\prime }}}=%
\frac{(2\pi )^{3}}{\Delta }\sum_{{\bf g}}\delta ((m^{\prime }-m)%
{\bf k}_{s}-{\bf g}).  \label{sumdelta1}
\end{equation}
Firstly let us show that in the sum over $m$ the main contribution
comes from the terms for which
$m{\mathbf{k}}_s{\mathbf{u}}_0\lesssim
{\mathbf{g}}{\mathbf{u}}_0$, or equivalently $m\lesssim \lambda
_s/a$, where $\lambda _{s}=2\pi /k_{s}$ is the wavelength of the
external excitation, and $a$ is the lattice constant. Indeed, due
to the $\delta $-function in  (\ref{sumdelta}) the Bessel function
enters in the form
$J_m({\mathbf{g}}{\mathbf{u}}_0-m{\mathbf{k}}_s{\mathbf{u}}_0)$.
For terms with $m{\mathbf{k}}_s{\mathbf{u}}_0 \gg
{\mathbf{g}}{\mathbf{u}}_0$, by taking into account that for
practically important cases $u_0/\lambda _s\ll 1$, and making use
of the asymptotic formula for the Bessel function, for large
values of the order we can see that
$J_m(m{\mathbf{k}}_s{\mathbf{u}}_0)\sim (2\pi m)^{-1/2}(\pi
eu_0/\lambda _s)^m$, $m\gg \lambda _s/a \gg 1$, and the
contribution of these terms is exponentially suppressed. This
conclusion is also valid for the sum over $m'$. Now let us
consider the contribution of the summands $m\neq m^{\prime }$
in the sum (\ref{SmSm}). It follows from formula (\ref{sumdelta1}) that $%
\left| m-m^{\prime }\right| \gtrsim \lambda _{s}/a$.  As $\lambda
_{s}/a\gg 1$, from here we see that one of the numbers $m$,
$m^{\prime }$ is large. Let it is the number $m^{\prime }$,
$m^{\prime }\gtrsim \lambda _{s}/a$. For the argument of the
Bessel functions one has the estimate ${\bf qu}_{0}\sim {\bf
gu}_{0}\sim 2\pi u_{0}/a$, and, hence, the ratio of the Bessel
function order to the argument is estimated as $ m^{\prime }/{\bf
qu}_{0}\sim \lambda _{s}/2\pi u_{0}$. As for the
practically important cases $u_{0}/\lambda _{s}\ll 1$ , for the function $%
J_{m^{\prime }}({\bf qu}_{0})$ we have
\begin{equation}
J_{m^{\prime }}({\bf qu}_{0})\sim \sqrt{\frac{a}{2\pi \lambda
_{s}}}\left( \frac{\pi eu_{0}}{\lambda _{s}}\right) ^{\lambda
_{s}/a}  \label{condm}
\end{equation}
It follows from here that under the condition $u_{0}/\lambda _{s}\ll 1$ the
contribution of the terms with $m\neq m^{\prime }$ in the sum (\ref{SmSm}%
) is small compared to the diagonal terms. In the case $m=m^{\prime }$ the
sum in the left hand side of (\ref{sumdelta1}) is equal to the number of
cells, $N$, in a crystal and the square of the modulus for the sum on the
left of Eq.(\ref{SmSm}) can be written as
\begin{equation}
\left| \sum_{n}e^{i{\bf qr}_{n0}}\right| ^{2}=N\frac{(2\pi
)^{3}}{\Delta }\sum_{m=-\infty }^{+\infty
}J_{m}^{2}({\bf qu}_{0})\left| S({\bf q}%
_{m})\right| ^{2}\sum_{{\bf g}}\delta ({\bf q}_{m}-{\bf g}).  \label{SmSmnew}
\end{equation}
Note that in this case we have no dependence on the phase $\varphi _{0}$.

In formula (\ref{sig2}) the first two terms in figure braces do
not depend on the direction of the vector ${\bf q}$ and correspond
to the contribution of incoherent effects. The third summand
depends on the orientation of crystal axes with respect to the
vector ${\bf q}$ and determines the contribution of coherent
effects. The corresponding part of the cross-section is known as
an interference term and, by taking into account the formula for
$\sigma _{0}({\bf q})$ (see, e.g., \cite{TerMik,Shulga}), can be
written as
\begin{equation}
\sigma _{c}=\frac{e^{2}}{(2\pi )^3 \omega ^{2}}\frac{%
q_{\perp }^{2}}{q_{\parallel }^{2}}\left| u_{{\bf q}}\right|
^{2}\left(
\frac{\omega \delta }{m_{e}^{2}}-1+\frac{2\delta }{q_{\parallel }}-\frac{%
2\delta ^{2}}{q_{\parallel }^{2}}\right) e^{-q^{2}\overline{u_{t}^{2}}%
}\left| \sum_{n}e^{i{\bf qr}_{n0}}\right| ^{2},  \label{sigcoh}
\end{equation}
where $u_{ {\bf q}}$ is the Fourier-transform of the atomic
potential, ${\bf q} _{\parallel }$ and ${\bf q}_{\perp }$ are the
parallel and perpendicular components of the vector ${\bf q}$ with
respect to the direction of the photon momentum ${\mathbf{k}}$,
$\omega $ is the photon frequency, $\delta =1/l_{c}$ is the
minimum longitudinal momentum transfer, and
$l_{c}=2E_{+}E_{-}/(\omega m_{e}^{2})$ is the formation length for
the pair creation process. Usually one writes the quantity
$u_{{\bf q}}$ as $4\pi Ze^2\left[ 1-F(q) \right] /q^{2}$, where
$Z$ is the number of electrons in an atom, and $F(q)$ is the
atomic form-factor. For the exponential screening of the atomic
potential one has $u_{{\bf q}}=4\pi Ze^2/(q^{2}+R^{-2})$ , with
$R$ being the screening radius.

The general expression for the pair creation cross-section can be
presented in the form
\begin{equation}\label{gencross}
d\sigma =N_0(d\sigma _n+d\sigma _c),
\end{equation}
where $d\sigma _n$ and $d\sigma _c$ are the cross-sections for the
non-coherent and coherent processes of the electron-positron pair
creation in a crystal per single atom. Using formulae
(\ref{SmSmnew}) and (\ref{sigcoh}) and integrating over ${\bf q}$
the cross-section for the coherent part can be presented as
\begin{eqnarray}
\frac{d\sigma _{c}}{dE_{+}} &=&\frac{e^{2}N}{\omega ^{2}N_0\Delta
}\sum_{m,{\bf g}}\frac{ g_{m\perp }^{2}}{g_{m\parallel
}^{2}}\left[ \frac{\omega ^{2}}{2E_{+}E_{-}} -1+2\frac{\delta
}{g_{m\parallel }}\left( 1-\frac{\delta }{g_{m\parallel }}
\right) \right] \left| u_{{\bf g}_{m}}\right| ^{2}\times   \label{dsigpm} \\
&&\times J_{m}^{2}({\bf g}_{m}{\bf u}_{0})\left| S({\bf g})\right|
^{2}e^{-g_{m}^{2}\overline{u_{t}^{2}}},\quad {\bf g}_{m}={\bf
g}-m{\bf k}_{s},  \nonumber
\end{eqnarray}
where the summation goes under the constraint
\begin{equation}
g_{m\parallel }\geq \delta .  \label{sumcond}
\end{equation}
For the simplest crystal with one atom in the elementary cell one
has $N=N_{0}$ and $S({\mathbf{g}})=1$. Formula (\ref{dsigpm})
differs from the corresponding formula for the creation of pairs
in an undeformed crystal (corresponding to the summand with $m=0$
and $u_{0}=0$, see, for instance, \cite{TerMik,Shulga}) by
replacement ${\bf g}\rightarrow {\bf g}_{m}$, and additional
summation over $m$ with weights $J_{m}^{2}({\bf g}_{m}{\bf
u}_{0})$. This corresponds to the presence of an additional one
dimensional lattice with the reciprocal lattice vector
$m{\mathbf{k}}_s$, $m=0,\pm 1,\pm 2,\ldots $. Note that by taking
into account the $\delta $-function in Eq. (\ref{SmSmnew}) the
momentum conservation law can be written down as
\begin{equation}
{\bf k}={\bf p}_{+}+{\bf p}_{-}+{\bf g}-m{\bf k}_{s},
\label{conslaw}
\end{equation}
where $-m{\mathbf{k}}_s$ stands for the momentum transfer to the
external field.

\section{Analysis of the general formula }
\label{sec3:an}

In formula (\ref{dsigpm}) for the pair creation cross-section the
main contribution comes from the terms with $g_{m\parallel }\sim
\delta $. It follows from here that the external excitation with a
wave vector ${\bf k}_{s}$ will influence on the process of pair
creation if $mk_{s\parallel }\gtrsim \delta $. As a consequence of
the well-known properties of the Bessel function, in the sum over
$m$ the main contribution is due to the summands with
\begin{equation}
m\lesssim {\bf g}_{m}{\bf u}_{0}\sim gu_{0}\sim \frac{2\pi
u_{0}}{a}. \label{mcond}
\end{equation}
From the last two relations it follows that it is necessary to take into
account the influence of external fields on pair creation by photons if
\begin{equation}
\frac{u_{0}}{\lambda _{s}}\gtrsim \frac{a}{(2\pi )^{2}l_{c}}.
\label{u0cond}
\end{equation}
It should be noted that at high energies $a/l_{c}\ll 1$ and
condition (\ref{u0cond}) does not contradict to the condition
$u_{0}/\lambda _{s}\ll 1$ . In view of the expression for the
formation length, condition (\ref{u0cond}) can be also written as
\begin{equation}
\frac{u_{0}}{\lambda _{s}}\gtrsim \frac{am_{e}}{8\pi
^2}\frac{m_e\omega }{E_{+}E_{-}}. \label{u0cond1}
\end{equation}
As for the process of pair creation $\omega >E_{\pm }$, this
condition is stronger than in the case of bremsstrahlung (see
\cite{Sahbrem}).

Let us consider the case of the simplest crystal with one atom in
the elementary cell assuming that the photon enters into the
crystal at small angle $\theta $ with respect to the
crystallographic axis $z$ of the orthogonal lattice. The
corresponding reciprocal lattice vector components are $g_i=2\pi
n_i/a_i$, $n_i=0,\pm 1,\pm 2,\ldots $, where $a_i$, $i=1,2,3$ are
the lattice constants in the corresponding directions. We can
write
\begin{equation}
g_{m\parallel }= g_{mz}\cos \theta +\left( g_{my}\cos \alpha
+g_{mx}\sin \alpha \right) \sin \theta ,  \label{gmpar}
\end{equation}
where $\alpha $ is the angle between the projection of the vector
${\bf k}$ on the plane $(x,y)$ and axis $y$. For small angles
$\theta $ the main contribution into the cross-section comes from
the summands with $g_{z}=0$ and we receive
\begin{equation}\label{crossc1}
\frac{d\sigma _c}{dE_{+}}\approx \frac{e^2}{\omega ^2\Delta }\sum
_{m,g_x,g_y}\frac{g_{\perp }^2}{g_{m\parallel }^2}\left[
\frac{\omega ^2 }{2E_{+}E_{-}}-1+2\frac{\delta
}{g_{m\parallel}}\left( 1-\frac{\delta }{g_{m\parallel}} \right)
\right] |u_{g_{m}}|^2J_{m}^{2}({\mathbf{g}}_m{\mathbf{u}}_0),
\end{equation}
where $g_{\perp }^2=g_{x}^{2}+g_{y}^{2}$, and the summation goes
over the region $g_{m\parallel}\geq \delta $ with
\begin{equation}
g_{m\parallel }\approx -mk_{z}+\left( g_{y}\cos \alpha +g_{x}\sin
\alpha \right) \theta .  \label{gmparmain}
\end{equation}
Note that in the argument of the Bessel function
${\mathbf{g}}_m{\mathbf{u}}_0\approx {\mathbf{g}}_{\perp
}{\mathbf{u}}_0$. It follows from here that if the displacements
of the atoms in the acoustic wave are parallel to the axis $z$
then the main contribution into the cross-section is due to the
summand with $m=0$ and the influence of the acoustic wave is
small. The most promising case is the transversal acoustic wave
propagating along the $z$ - direction. If the photon moves far
from the crystallographic plane (the angles $\alpha $ and $\pi
/2-\alpha $ are not small) the expression under the sum is a
smooth function on $g_x$ and $g_y$, and the summation over these
variables can be replaced by integration:
\begin{equation}\label{sumtoint}
\sum _{g_x,g_y}\to \frac{a_1a_2}{(2\pi )^2}\int dg_xdg_y.
\end{equation}

\begin{figure}[t]
\begin{center}
\begin{tabular}{ccc}
\epsfig{figure=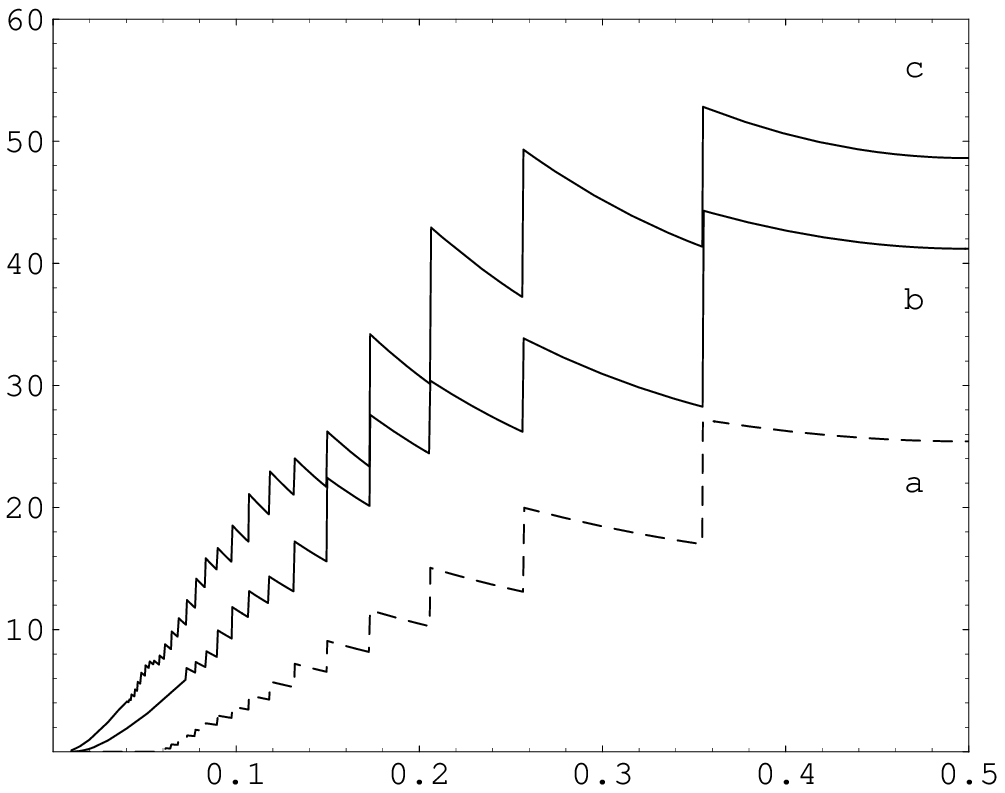,width=6cm,height=6cm} & \hspace*{0.5cm} & %
\epsfig{figure=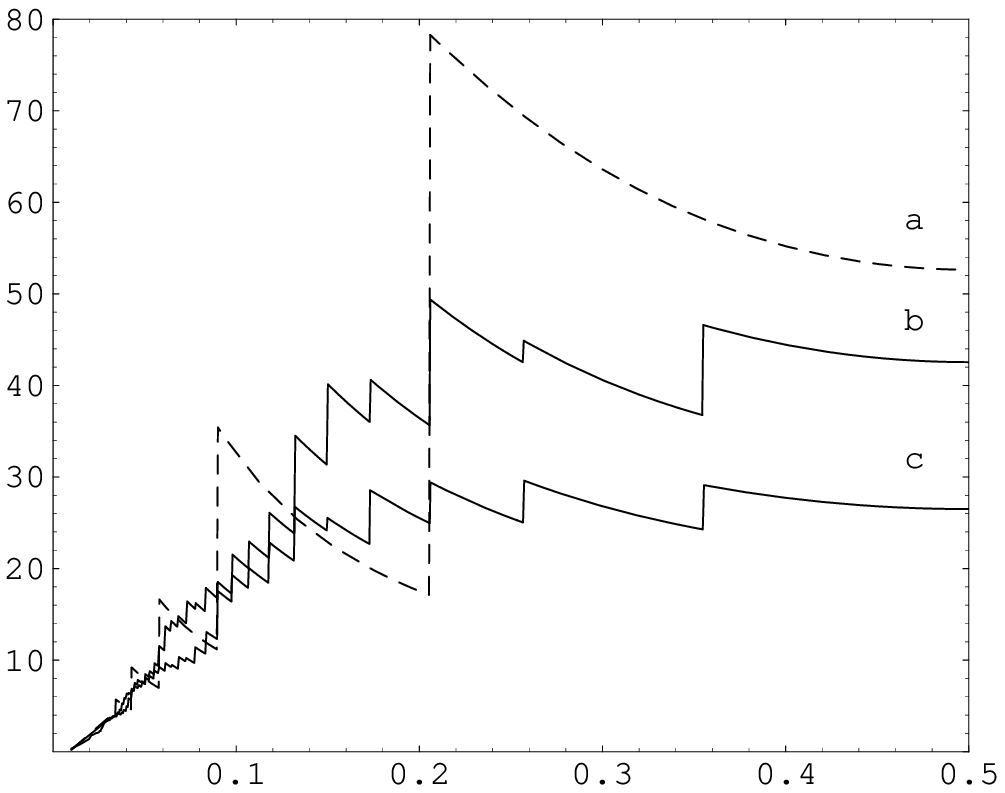,width=6cm,height=6cm}
\end{tabular}
\end{center}
\caption{Pair creation cross-section, $(m_e^2\omega
/Z^2e^6)d\sigma _c/dE_{+} $, as a function of $E_{+}/\omega $ for
various values of the hypersound amplitude: $2\protect\pi u_0/a_1=
0$ (a, dashed curves), 4 (b), 8 (c) and for $\psi =0.0005$ (left),
$\psi =0.0035$ (right). The values for the other parameters are as
follows: $\protect\omega =100$ GeV, $a_1/\protect\lambda _s=5\cdot
10^{-4} $, $a_1/2\protect\pi R=1$.} \label{fig1psi}
\end{figure}
We now assume that the photon enters into the crystal at small
angle $\theta $ with respect to the crystallographic axis $z$ and
near the crystallographic plane $(y,z)$ ($\alpha $ is small). In
this case with an increase of $\delta $ some sets of terms in the
sum will fall out. This can essentially change the cross-section.
Two cases have to be distinguished. Under the condition $\delta
\sim 2 \pi \theta /a_2$ in Eq. (\ref{crossc1}) for the
longitudinal component one has
\begin{equation}\label{condgy}
  g_{m\parallel }\approx
-mk_{z}+\theta g_{y}\geq \delta .
\end{equation}
This relation does not depend on the component $g_x$ and the
summation over this component can be replaced by integration $\sum
_{g_x}\to (2\pi /a_1)\int dg_x $. When ${\mathbf{u}}_0\parallel
{\mathbf{a}}_1$, for exponential screening the corresponding
integral is expressed in terms of the hypergeometric functions.
Here we will consider in detail the simpler case
${\mathbf{u}}_0\parallel {\mathbf{a}}_2$. For the exponential
screening after the elementary integration over $g_x$ we obtain
\begin{equation}\label{farplane}
  \frac{d\sigma _c}{dE_{+}}\approx \frac{4\pi ^2Z^2e^4}{\omega ^2a_2a_3 }\sum
_{m,g_y}\frac{2g_{y}^2+R^{-2}}{g_{m\parallel}^2(g_y^2+R^{-2})^{3/2}}\left[
\frac{\omega ^2 }{2E_{+}E_{-}}-1+2\frac{\delta
}{g_{m\parallel}}\left( 1-\frac{\delta }{g_{m\parallel}} \right)
\right] J_{m}^{2}(g_yu_{0}),
\end{equation}
where the summation goes under the condition (\ref{condgy}).

Now we will assume that $\delta\sim 2 \pi \theta \alpha /a_1$. The
main contribution into the sum in Eq. (\ref{crossc1}) is due to
the terms with $g_y=0$ and we have two sums: over $m$ and $n_1$,
$g_x=2\pi n_1/a_1$. For the corresponding cross-section one
receives:
\begin{equation}\label{crossc2}
\frac{d\sigma _c}{dE_{+}}\approx \frac{e^2}{\omega ^2\Delta }\sum
_{m,n_1}\frac{g_{x}^2}{g_{m\parallel}^2}\left[ \frac{\omega ^2
}{2E_{+}E_{-}}-1+2\frac{\delta }{g_{m\parallel}}\left(
1-\frac{\delta }{g_{m\parallel}} \right) \right]
|u_{g_{m}}|^2J_{m}^{2}(g_xu_{0x}),
\end{equation}
where
\begin{equation}\label{gmperpc2}
g_{m\parallel }\approx -mk_z+g_x\psi ,\quad \psi =\alpha \theta ,
\end{equation}
and summation goes over the values $m$ and $n_1$ satisfying the
condition
\begin{equation}
\left| n_1\psi -ma_{1}/\lambda _{s}\right| \geq
\frac{m_{e}^{2}a_{1}}{4\pi E_{+}(1-E_{+}/\omega )}.
\label{sumcond2}
\end{equation}
\begin{figure}[t]
\begin{center}
\begin{tabular}{c}
\psfig{figure=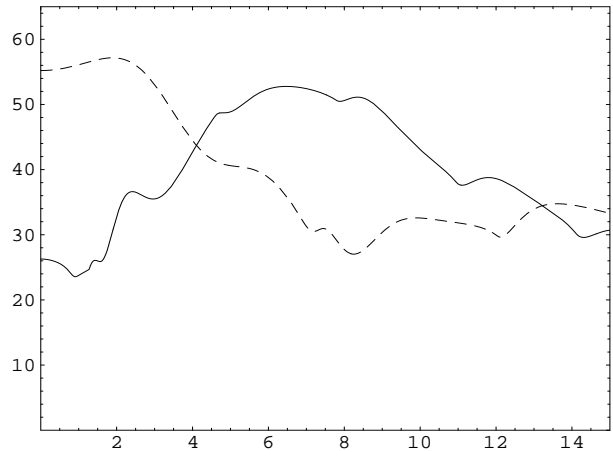,width=8cm,height=6cm}
\end{tabular}
\end{center}
\caption{Pair creation cross-section, $(m_e^2\omega
/Z^2e^6)d\sigma _c/dE_{+} $, as a function of $ 2\pi u_0/a_1$ for
the positron energy corresponding to $E_{+}/\omega =0.4$ and for
$\psi =0.0005$ (full curve), $\psi =0.0035$ (dashed curve). The
values for the other parameters are the same as in fig.
\ref{fig1psi}.} \label{fig2u0}
\end{figure}
In this case the most favorable conditions to have an influence on
the pair creation cross-section due to external fields are ${\bf
u}_{0}\parallel {\bf a}_{1}$ (to have large values for $m$) and
${\bf k}_{s}\parallel {\bf a}_{3}$ (to have large values for
$mk_{z}$). We have numerically evaluated the pair creation
cross-section by making use of formula (\ref{crossc2}) for various
values of parameters $\psi $, $u_0$, $\lambda _s$. The
corresponding results show that, in dependence of these
parameters, the external excitation can either enhance or reduce
the cross-section. As an illustration in fig. \ref{fig1psi} we
have depicted the quantity $(m_e^2\omega /Z^2e^6)d\sigma _c/dE_{+}
$ as a function of $E_{+}/\omega $ in the case of cubic lattice
($a_1=a_2=a_3$) and exponential screening of the atomic potential
for $u_{0}=0$ (dashed curves (a)), $2\pi u_{0}/a_{1}=4$ (curves
(b)), $2\pi u_{0}/a_{1}=8$ (curves (c)) and for $\psi =0.0005$
(left), $\psi =0.0035$ (right). The values for the other
parameters are as follows: $\omega =100$ GeV, $a_{1}/\lambda
_{s}=5\cdot 10^{-4}$, $a_{1}/2\pi R=1 $. As the cross-section is
symmetric under the replacement $E_{+}/\omega \to 1-E_{+}/\omega $
we have plotted the graphs for the region $0\leq E_{+}/\omega \leq
0.5$ only. In fig. \ref{fig2u0} we have presented the cross
section evaluated by Eq. (\ref{crossc2}) as a function of $2\pi
u_0/a_1$ for the positron energy corresponding to $E_{+}/\omega
=0.4$ and for $\psi =0.0005$ (full curve), $\psi =0.0035$ (dashed
curve). The values for the other parameters are the same as in
fig. \ref{fig1psi}.

\section{Conclusion} \label{sec4:conc}

In this paper we have investigated the electron-positron coherent
pair creation by high-energy photons in a single crystal in the
presence of a hypersonic wave. If the displacements of the atoms
in the crystal under the influence of hypersound have the form
(\ref{uacust}), the coherent part of the corresponding
cross-section per single atom, averaged on thermal fluctuations,
is given by formula (\ref{dsigpm}). To compared with the
cross-section in an undeformed crystal this formula contains an
additional summation over the reciprocal lattice vector
$m{\mathbf{k}}_s$ of the one dimensional superlattice induced by
the hypersonic wave. The contribution for a given $m$ is weighted
by $J_m^2({\mathbf{g}}_m{\mathbf{u}}_0)$, where the vector
${\mathbf{g}}_m$ is defined as in Eq. (\ref{dsigpm}). We have
substantiated that the influence of the hypersound on the
cross-section can be remarkable under the condition
(\ref{u0cond1}). Note that for $u_0\gtrsim a$ this condition is
less restrictive than the naively expected one $l_c\gtrsim \lambda
_s $. In section \ref{sec3:an} we have considered the most
interesting case when the photon enters into the crystal at small
angle with respect to a crystallographic axis (axis $z$ in our
consideration). The main contribution into the coherent part of
the cross-section comes from the crystallographic planes, parallel
to the chosen axis. The behaviour of this cross-section as a
function on the positron energy essentially depends on the angle
between the projection of the photon momentum on the plane $(x,y)$
and a crystallographic plane. The corresponding numerical
evaluations show that the presence of the hypersonic wave can
either enhance or reduce the cross-section. This can be used to
control the parameters of the positron sources for storage rings
and colliders. Note that for the positron sources based on the
channeling radiation the hypersound can also substantially
influence on the corresponding intensity of the radiated photons
(see Refs. \cite{Mkrtch1}).

\section*{Acknowledgment}

The work has been supported by Grant no. 1361 from Ministry of
Education and Science of the Republic of Armenia.

\end{document}